# AUTONOMOUS TOOLS FOR GRID MANAGEMENT, MONITORING AND OPTIMIZATION

## DESIGN AND DEVELOPMENT PROPOSAL

Wojciech Wiślicki

Interdisciplinary Centre for Mathematical and
Computational Modelling
University of Warsaw

**Abstract**

We outline design and lines of development of autonomous tools for the computing Grid management, monitoring and optimization. The management is proposed to be based on the notion of utility. Grid optimization is considered to be application-oriented. A generic Grid simulator is proposed as an optimization tool for Grid structure and functionality.

# 1. An autonomous tool for Grid monitoring and management

**General description**

This task aims to invent metrics useful for management of a distributed networking system, in particular a computational Grid. The metrices have to be proved efficiently calculable, based on tools for Grid monitoring, and easily interpretable in terms of managerial decisions. A management and monitoring tool based on these metrics can be either used in the offline mode or be further implemented as autonomous functions in logical devices like Resource Broker or Virtual Organization Manager.

Management of the Grid requires:
1. Efficient monitoring tools for collecting information on the current state and history of Grid resources, i.e. loads for different Grid logical components (computing elements, resource brokers, information databases, monitoring tools, storage elements, routers) and used bandwidths on local and external links. All these measures have to be compared to available resources.
2. A limited number of characteristics for relevant features of the Grid which can be easily used by either a human or an autonomous system. These characteristics should be intuitive and easily interpretable in terms of managerial decisions. For example, if an intensive collective flow of large packages appears in a particular segment of the Grid, it has to be easily detectable in order to adjust maximum transition units on switches or allow for higher information throughput.

Grid management is also related to accounting which, in addition to measurements of resource usage, requires good timing and quality measures. Metrics relevant for accounting should represent combinations of used resources, response times to user's requests (which defines degree of interactivity of applications), and overall time spent by a job in the system, reliability of the system (breaks, failures, stability, etc.).

We propose an approach based on the notion of utility[1] and

---
[1] Among large variety of literature where the notion of utility is discussed, one may either consult a classical opus by J. von Neumann and O. Morgenstern, Theory of

description of the Grid as a stochastic system close to equilibrium, i.e. with utility uniformly distributed over all degrees of freedom of the system. In practical terms this requires that any large-scale variation of the system happens with a characteristic time much longer than local utility flow between system's components. However, such approach does not have to appear neither the most efficient computationally nor sensitive to all important features of the Grid.

The question of whether working conditions of the Grid are in the vicinity of equilibrium, or non-equilibrium quasi-stationary state, is an open issue on which, up to our knowledge, no satisfactory study was reported in literature so far. In order to treat this appropriately, one has to perform some theoretical work on the notion of subsystem for the Grid, recognizing its degrees of freedom, and carefully checking working conditions, both using data and simulations. Furthermore, we are planning to use a high-level Grid simulator, which may be developed in parallel as a second task of this Project, to investigate equilibration of the Grid.

Another important factor for Grid working conditions is given by pricing, i.e. defining a money-valued function of the utility. Pricing may serve as an additional tool for assuring stability and providing with control to the management agent. Studies of this subject are known[2]. Their relevance for our purposes needs to be evaluated.

**Preliminary steps**

At this stage we aim to:
1. Investigate monitoring methods and tools used in existing Grid systems and projects, as e.g. those based on GLOBUS[3], UNICORE[4], LEGION[5], etc. Special attention has to be devoted to incorporating monitoring libraries, as e.g. GANGLIA[6], in monitoring middleware currently in use in large Grid projects, as GRID-Ice[7] or GOC Grid

---

   

Monitoring[8] in EGEE[9] project.
2. Study existing definitions of utility function for telecommunication and computer systems, used e.g. for solving scheduling problems[10], economic problems related to accounting in computing networks. Since our ultimate goal is to invent or adopt utility suitable for optimization of such complex system like computing Grid, some more general study of utility has to be performed, viz. finding relevant independent variables, efficient parameterizations and, ideally, analytic expressions for them. Interesting and useful thread in such investigation would be adoption of efficient estimation of utility known in economy as the conjoint method.
3. Investigate existing predictor and forecaster tools.
4. Investigate role of an accounting system for equilibration of the Grid. It seems intuitively possible that accounting for resource usage should speedup equipartition of utility among all degrees of freedom of the Grid, thus improving stationarity of its working conditions. Currently, an example of an accounting system, DGAS[11], provided by gLite middleware in the framework of EGEE and LCG projects, is under evaluation[12]. DGAS collects information about usage of Grid resources by users and groups of users (including VO). This information can be used to generate reports and billings, and also to implement resource quota system.
5. Investigate, and possibly conclude, on the impact of pricing and quality-of-service on equilibration of the Grid. Results of this subtask are important for the second task, because finding equilibrium is normally related to the optimization of the system.

Important in these studies is completeness of knowledge gained, in order to avoid reinventions and ensure the best possible quality of own developments.

**Specification of requirements for the tool**

One has to define synthetic metrics of the system, based on data collected by monitoring tools and utilities evaluated from them. Examples of such quantities may be entropies or generalized Renyi entropies[13] which should remain constant for uniformly loaded

---

[8] http://goc.grid-support.ac.uk/gridsite/monitoring/
[9] http://egee-intranet.web.cern.ch/egee-intranet/gateway.html
[10] D. Vengerov, A reinforcement learning framework for utility-based scheduling in resource-constrained systems, SMLI-TR 2005-141
[11] User's Guide for DGAS Services, https://edms.cern.ch/document/571271
[12] LHC Computing Grid Technical Design Report, LCG-TDR-001, CERN-LHCC-2005-024,
[13] Cf. e.g. C. Beck, Physica D41(1990)67
A. Majka and W. Wislicki, Physica A322(2003)313

network and exhibit a rapid decrease in case of collective behaviours reducing number of degrees of freedom in the system. Another examples are fluctuations of agent's number in a subsystem or first moments of utility[14]

One has to decide, however, of whether quantities defined using thermodynamic analogies, with possible upgrades to non-extensive thermodynamics in case of non-additive utilities, are efficient and sensitive to important features of the system. If not, these have to be complemented by more appropriate measures.

Effective computability of such measure represents an important requirement. For networks of complex topology, computation of a complete function of state is often a hard combinatorial problem. This is obviously unacceptable for management and some algorithmic shortcuts have to be found then. For example, in case of a large number of states, those with very low utilities have to be cutoff and a possible bias due to that procedure has to be estimated.

Since validity of the equilibrium assumption for the system is not a priori obvious, this has to be clearly stated as a possible limitation for interactive Grids. However, progressing technologies make response times fast enough that even in such case it may represent acceptable approximation. Validity of this assumption can be independently checked using the same monitoring tools.

**Design of the management system**

1. Provided appropriate measures are found, either utility-based or modified, they have to be computed for existing Grids. To this end, one has to parameterize, or model directly using Monte Carlo methods, basic components of the Grid in a relevant variables' space. For example, if a total job time is considered, all time components as queuing time, transfer time, execution time etc. have to be properly represented in the utility function. This is not an easy task for devices like resource brokers, where number of internal operations is high and boundary conditions are varying (cf. Fig. 1).

---

[14] A. Majka and W.Wislicki, Physica A337(2004)345

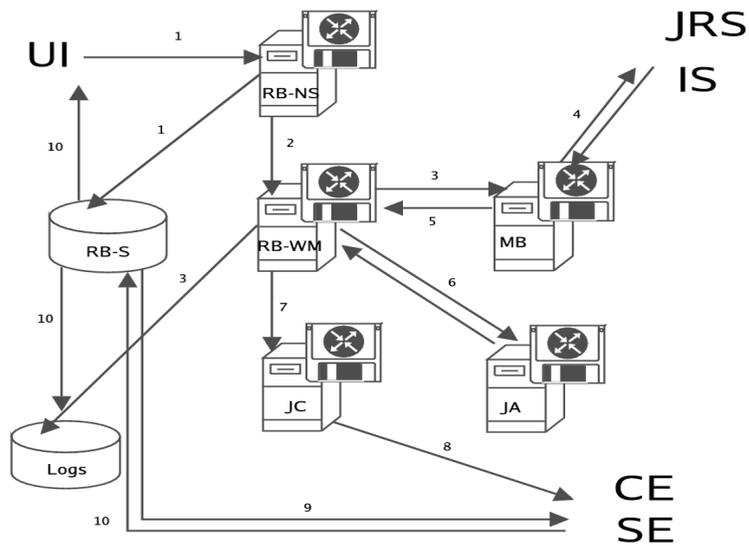

Fig. 1 *Structure and functionality of the Resource Broker (RB) used in Grids based on Globus Toolkits. The abbreviations stand for the following components of the Broker: User Interface (UI), Job Replica System (JRS), Information System (IS), Computing Element (CE), Storage Element (SE), Networking System (NS), Workload Manager (WM), Job Collector (JC), Job Adapter (JA), Matchmaker Box (MB). Numbers show the workflow in the system.*

Also, proper modelling of access to logical site components (cf. fig. 2) like computing element, storage element, and also transfer times between user interfaces and other machines, computing times in computing elements, has to be performed.

2. Design of the system needs to distinguish at least two cases
    - the user-oriented approach,
    - the resource-oriented approach.

In the first case, jobs may be treated as agents competing for resources, whereas in the second one resources compete for users. For Grids with implemented accounting system, the second approach seems to be more appropriate.

3. Another distinction has to be made between choice models and games. If agents in user-oriented approach strictly define resources (which can be done in Globus Toolkit using Job Description Language) with no freedom left for Resource Broker, the system behaves like a game and agents looking for best resources are likely acting as in the minority game.

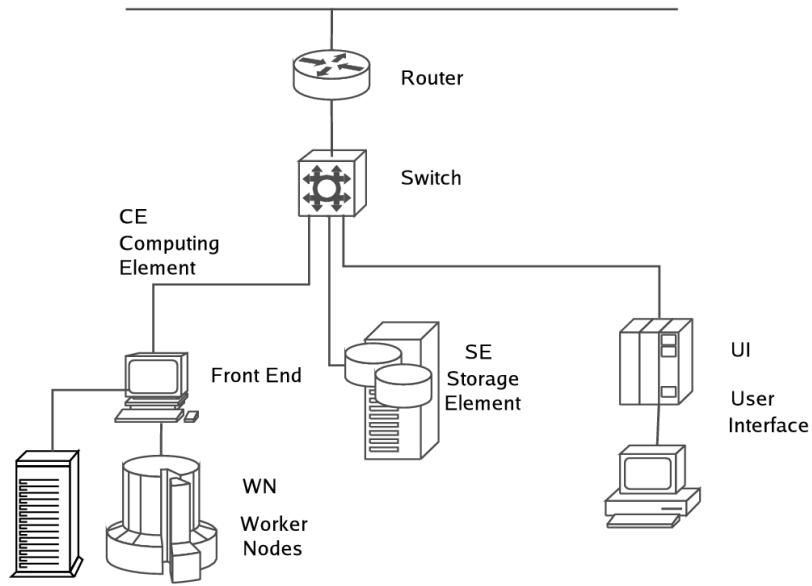

Fig. 2 Example of the Grid site with its basic components: the User Interface (UI), Computing Element (CE) with the Front End (FE) and Worker Nodes (WNs), and Storage Element (SE).

Contrary to this, if unconstrained freedom is left to the Resource Broker, we are closer to the choice model, where the Resource Broker is a principal decision maker in the system. Resource Brokers with good access to information about the system act similarly to those on a stock market where decisions are taken using money as utility functions.

# 1. The application-oriented simulator for autonomous Grid optimization and scheduling

**General description**

Being very complex systems with numbers of states growing factorialy with numbers of nodes, links and jobs, computational Grids need extensive simulations for efficient handling and management. Optimization and management problems, however, if formulated in general terms, cannot be treated without specifying applications one wants to work with, as it may happen that different applications may lead to contradictory requirements to the system. Therefore, optimization may become application-oriented, which means that application itself has to be accounted for when constructing the model of the Grid. Since a high degree of autonomy is required in today's management systems for telecommunication networks, proper modelling and simulation of its autonomic functions is called for. In case of Grids, such central autonomic functional unit may be e.g. a Resource Broker, playing the role of the central manager of the Grid, evaluating jobs, consulting databases and information systems, matchmaking jobs with resources, preparing jobs for execution and closely collaborating with schedulers and forecasters[15]. Another example, to our knowledge not yet existing, would be an automatic manager of Virtual Organizations, doing dynamic resource allocations, or even adjusting topology (connectivity scheme) to current needs.

Real applications and logical machines, however, are too complex to work with when simulating the Grid. In order to overcome this, one may use a real Grid for measurements of interesting characteristics of a real application, e.g. how much of resources it consumes, what are its times of transfers, computations, usage of databases, queuing times in different configurations etc. Based on these results, one may construct a model application, much simpler in functionality than the original one, but exhibiting all original's features.

Once the model application is defined, the next step should be a construction of a simplified model of the Grid building blocks, as Computing Elements, including front-end machines and Worker Nodes, User Interfaces, Storage Elements, switches, routers, other logical machines, information indices and other databases, etc. Although ideally one could perform a complete Monte Carlo simulation of these components, that may appear too complex and time consuming, mainly due to variety of boundary conditions. But effective parametrizations of these devices can be done and used for that purpose.

---

[15] An example of a forecaster prepared by K. Nawrocki, A. Padee and K. Wawrzyniak in the framework of the EU-supported CrossGrid Project for Grid based on Globus Toolkit, see http://grid.fuw.edu.pl/cgi-bin/gmdat/engine/grid2.pl

We can formulate four principal subjects:

3. Our maximalistic purpose is to work-out a tool for tuning of the Resource Broker, in order to make it optimal with respect to given application or family of applications. This tool can be based on a Grid simulator.

3. The question, raised already in the first task, described in chapter 1, and concerning validity of the (quasi-)stationarity assumption for working condition of the Grid, can be efficiently addressed using Grid simulation tool we pretend to design in present task. Complete treatment of this problem cannot be done before full implementation of the simulator is done. However, for the particular purpose of investigating equilibration and optimization, a simplified and more generic Grid simulator can be used. Using it we hope to conclude, partially in the framework of task 1, on the following issues:
    1. Is it realistic to assume that computing Grid works in stationary regime and what are the minimum conditions for that?
    2. Is it profitable that such stationary conditions are found, and for whom (grid management profit, overall users' community profit, profits for specific user groups, efficiency of management?)
    3. What is the role of accounting system for Grid's optimization? Do existing accountings (cf. e.g. ref.[16]) help with that respect?
    4. What is the impact of the quality-of-service and pricing on Grid optimization technique?
    5. What is the role of an advanced resource reservations functionality, and also benchmark-based resource selections, considered for some resource brokers [17]

4. The third goal is a design of an automatic redefinition of resources

---

[16] User's Guide for DGAS Services, https://edms.cern.ch/document/571271 and LHC Computing Grid Technical Design Report, LCG-TDR-001, CERN-LHCC-2005-024

[17] E. Elmroth and J. Tordsson, www.cs.umu.se/~elmroth/ papers/ elmroth_tordsson_para04.pdf and Advanced Parallel Computing, to appear in Lecture Note in Computer Science, Springer.

on the Grid (used in Virtual Organization management systems), including its topological structure (or connectivity scheme), allocation of storage and CPUs to Virtual Organizations, locations of data replicas etc., according to recommendations from the Grid simulator. For example, for performance of data-intensive jobs, topology of connections between computing elements and storage elements is crucial (cf. fig. 3). At the moment, we are not aware of any good and autonomic management system with such functionality, e.g. the Virtual Organization Membership Service (VOMS)[18], currently in use in the EGEE Project[19], offers only management of users databases.

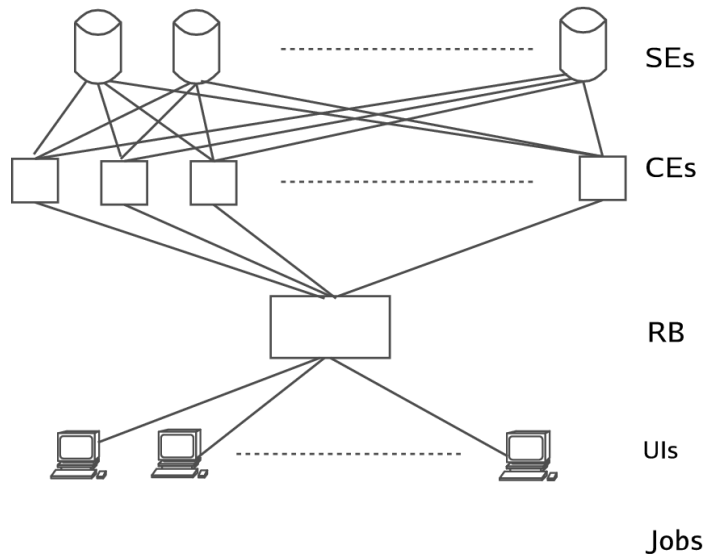

Fig. 3  Topological structure of the the Grid based on Globus Toolkit. The components are: Users' jobs (Jobs) submitted through User Interfaces (UIs), distributed by the Resource Broker (RB) over Computing Elements (CEs) and Storage Elements (SEs). Normally, a queuing system at each CE can manage local jobs with a number of predefined queues, possibly with user profiling[20]. It may appear profitable to send a job to the other CE, where the queue is shorter, and to transfer data to it through the network from a distant SE. Links between CEs and SEs are logical and do not necessarily reflect hardware topology of the network.

This approach is also very interesting from theoretical point of view in the theory of complex systems, because allowing different features of the Grid to be random variables, corresponds to different definitions of statistical ensembles, not having direct

---

[18] http://vdt.cs.wisc.edu/VOMS-documentation.html
[19] http://egee-intranet.web.cern.ch/egee-intranet/gateway.html
[20] An example of a user-profiling scheduler is MAUI queuing system, cf. http://www.nsc.liu.se/systems/cluster/grendel/maui.html

counterparts in natural systems [21].

3. Dynamic topology adjustment and management can be performed both at the level of network fabric or, using fixed existing Grid hardware layer, by redefinition of links and bandwidth in network devices (routers, bandwidth splitters). This can be done using existing algorithms in the framework of reinforcement learning method[22] with its interesting adaptations to minority games schemes.

**Grid-based self-management services**

Generally, such system can be considered independently of specific computing platforms and only approaching implementation one has to focus on existing solutions. However, different computing platforms can develop their specific middleware solutions. Those computing platforms should be investigated in order to find their software tools helpful for evaluation of applications, simulations of Grid components and their functionalities (both hardware and software), simulations of applications, building queuing systems and building schedulers and resource brokers and finally, redesigning dynamically Grid connectivity schemes and resource allocations. Results of this investigation may be important for both the final solution and choice of computing platform for design and implementations.

This research, if done completely, is rather time consuming because of necessary access to different computing platforms. Important part of such research, however, can be performed with a simulator at a generic level, with no complete account of specific Grid toolkits and computing platforms. This needs a two-level description of the simulator: (i) a generic one, where the basic functionality of Grid components is simulated for a set of logical machines and services common to representative Grid toolkits and middleware, and (ii) specific models for Grids, including simulations and parameterizations of platform- and middleware-dependent components.

In addition, existing Grid simulators should be recognized and investigated at this stage[23]. Our further developments have to account

---

[21] A. Majka and W. Wislicki, http://arxiv.org/pdf/cond-mat/0510699, Eur. Phys. J. B48(2005)271

[22] A. Galstyan and K. Czajkowski and K. Lerman, Resource Allocation in the Grid Using Reinforcement Learning, in Proceedings of the Third International Joint Conference on Autonomous Agents and Multiagent Systems, IEEE Computer Society, Washington, DC, USA, 3(2004)1314

[23] BeoSim, http://www.parl.clemson.edu/~wjones/research/
GridExplorer, http://www.lri.fr/~fci/GdX/

for current achievements and either supersede them or extend from them.

## Requirements for Grid services and functions for building autonomic service functions

At this stage one identifies Grid services and functions which can serve as building blocks for our simulator and perhaps also for final implementation of real resource brokers.

In order to do that one investigates, and possibly performs, modifications of existing resource brokers, resource allocation managers and routers, according to recommendations of the simulator.

Formulating requirements for new functionalities of the Grid, one has to account for two important extensions of Grid components:
1. massive data producers have to be treated by resource brokers as a special category of agents. Examples of such jobs are data acquisition systems or billing systems. These jobs can be either run as services or as clients. In the latter case they represent a specific type of a client with predefined resources and rather long execution time. They should be executed with no queuing.
2. In some cases, relationships between jobs cannot be ignored, giving rise to clusters of mutually dependent agents on the Grid. This special kind of constraints can be crucial for some applications and can be very different from typical constraints, depending on applications. For example, parallelized jobs using MPICH[24] may need predefined numbers of CPUs and bandwidth and some minimum number of data replicas. Typical pipelined jobs with sequential, chain-like IO dependencies, may save massive SE usage for intermediate storage but would be more CPU-intensive.

---

ChicSim, http://people.cs.uchicago.edu/~krangana/ChicSim.html
Optorsim, http://hpc.sagepub.com/cgi/content/
Gridsim, http://www.buyya.com/gridsim/
[24] http://www-unix.mcs.anl.gov/mpi/mpich/

**Design and implementation of novel Grid-based autonomic management systems**

It may appear that using Grid simulator built in this proposal, one finds a solution for resource brokerage or automatic resource- or topology designer not represented in any existing solution. This is supported by our overview of the literature, and both attendance and material studies of principal recent conferences and workshops. It is therefore desirable to study it further and design it, provided such solution is found. Amendments to existing resource brokers should be also documented, wherever possible.